\documentclass[aps,prl,preprint,groupedaddress]{revtex4}
\usepackage{amssymb}
\usepackage{amsmath}
\usepackage{graphicx}
\begin{document}

\title{Structures and energetics of hydrocarbon molecules in a wide hydrogen
chemical potential range}
\author{Y. X. Yao}
\affiliation{Ames Laboratory-U.S. DOE and Department of Physics and Astronomy, Iowa State
University, Ames, Iowa IA 50011}
\author{C. Rareshide}
\affiliation{Physics department, Stetson University, DeLand, FL 32723}
\author{T. L. Chan, C. Z. Wang, and K. M. Ho}
\affiliation{Ames Laboratory-U.S. DOE and Department of Physics and Astronomy, Iowa State
University, Ames, Iowa IA 50011 }

\begin{abstract}
We report a collection of lowest-energy structures of hydrocarbon molecules $%
C_{m}H_{n}$ (m=1-18; n=0-2m+2). The structures are examined within a wide
hydrogen chemical potential range. The genetic algorithm combined with
Brenner's empirical potential is applied for the search. The resultant
low-energy structures are further studied by \textit{ab initio }quantum
chemical calculations. The lowest-energy structures are presented with
several additional low-energy structures for comparison. The results are
expected to provide useful information for some unresolved astronomical
spectra and the nucleation of growth of nano-diamond film.
\end{abstract}

\maketitle

\section{Introduction}

Hydrocarbon molecules, especially polycyclic aromatic hydrocarbons (PAHs),
have received much attention for the broad astronomical interests. PAHs were
proposed to be the possible carriers of the unidentified infrared (IR)
emission bands\cite{Leger84}, the interstellar ultraviolet (UV) extinction
curve\cite{Joblin92,Clayton03} and the diffuse interstellar bands\cite%
{Zwet85,Tielens95,Salama96}. Though quite a few achievements have been made
in recent years\cite{Allamandola99,Allamandola03}, the PAH hypotheses are
not fully addressed. Possible different charged states of PAHs may increase
its complexity, however, one intrinsic difficulty could be stemmed from the
very complicated phase space of hydrocarbons due to the enormous bonding
ability of carbon. Furthermore, the phase space is also dependent of the
hydrogen chemical potential, as the observed IR spectra may be different in
different interstellar environments\cite{Allamandola03}.

In this paper, we focus on providing a collection of lowest-energy
structures for $C_{m}H_{n}$ (m=1-18; n=0-2m+2) in a wide hydrogen chemical
potential range with the aid of a series of unbiased global searches. Our
results are expected to be helpful for addressing the above PAH hypotheses.
Moreover, since the most abundant reactive elements in the cosmos are carbon
and hydrogen, our database should be helpful in elucidating the evolution of
carbon from its birthsite in circumstellar shells through interstellar
medium, which may be related to astrobiology ultimately\cite{Allamandola03}.
Useful information for a better understanding of the mechanism of nucleation
and growth for nano-diamond films may also be extracted from our results\cite%
{Lifshitz02,Michaelson05}.

\section{Computational Methods}

The global structure optimization is based on our early developed genetic
algorithm (GA)\cite{Deaven95}. An illustrative online version of the code
for the optimization of a two-dimensional map has been published in nanohub%
\cite{nanohub}. Each GA run is started with a randomly generated pool with
typically 100 structures which are relaxed to local minimum with Brenner's
empirical potential\cite{Brenner90}. The size of pool may vary with
different investigated systems. The random generation of the pool may be
guided by some physical intuition. The evolution of the pool is realized by
performing mating operations. Each operation includes randomly selecting two
structures (parents) from the pool, cutting them with a common plane
randomly selected, then combining the opposite parts relative to the cutting
plane from the parents to create a new structure (child). The child
structure is relaxed within Brenner's model\cite{Brenner90}. The decision of
replacing the highest-energy structure in the pool with the child structure
is based on energy criterion $f$ defined as%
\begin{equation}
f=(E-n_{H}\mu _{H})/n_{C}
\end{equation}%
where $E$ is the total energy of the cluster with $n_{H}$ hydrogen atoms and
$n_{C}$\ carbon atoms. $n_{C}$\ is fixed for each GA run, while $n_{H}$\ may
be varied in a physically reasonable range. $\mu _{H}$\ is the hydrogen
chemical potential, which may be fixed to some particular value(s)\cite%
{Chan06} or a range of interest. In this work, we are interested in a wide
hydrogen chemical potential range, which is taken as (-6eV, 0) in Brenner's
model, i.e., the lowest (highest)-hydrogen chemical potential value
guarantees that the optimum molecules are pure carbon clusters (alkanes). In
practice $\mu _{H}$\ is uniformly sampled in this chemical potential range
by step size of 0.05 eV. The mating operations performed with parents of
different number of hydrogen atoms may provide a superior sampling of the
potential energy landscape\cite{Chan06}. Each GA run has typically 10000
generations with 15 random pairs of molecules mated for each generation. At
the end of each GA run, the candidate structures are further relaxed by
\textit{ab initio} quantum chemistry method.

\begin{figure}[tbp]
\begin{center}
\includegraphics[width=8.00cm]{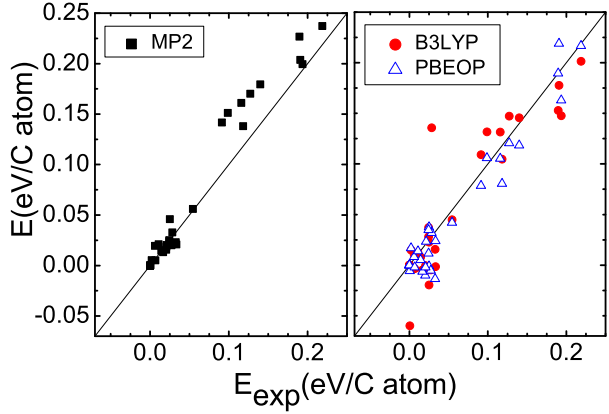}
\end{center}
\caption{
(Color online) Computed isomer energy differences at MP2 (black square), DFT-B3LYP (red
circle) and DFT-PBEOP (blue triangle) level versus experimental values for
34 hydrocarbon molecules. 
}
\label{ecorr}
\end{figure}

Recently it has been pointed out that density functional theory (DFT) is
unreliable for computing hydrocarbon isomer energy differences\cite%
{Wodrich06,Schreiner06}. Fig.\ref{ecorr} shows the correlation between
experimental isomer energy differences and those calculated at second order M%
\o ller--Plesset perturbation theory\ (MP2), DFT-B3LYP, DFT-PBEOP levels
with basis set of 6-31(d) for 9 $C_{6}H_{10}$, 7 $C_{6}H_{12}$, 5 $%
C_{6}H_{14}$, 4 $C_{7}H_{14}$ and 9 $C_{7}H_{16}$ molecules. MP2 gives a
best correlation with the experiments, especially for the low-energy
molecules. Hence all the candidate hydrocarbon structures in the final pool
from GA are relaxed at level of MP2/6-31G(d) using GAMESS package\cite%
{Schmidt93}.

\section{Results}

We systematically searched for the lowest energy structures of hydrocarbon
molecules $C_{m}H_{n}$ (m=1-18; n=0-2m+2) in a wide hydrogen chemical
potential range of (-6eV, 0). Fig.\ref{C6}-\ref{C18} show MP2-relaxed
lowest-energy structures from our unbiased global searches. Several
low-energy isomers were also shown for comparison. The ground state
geometries of pure carbon clusters have been extensively studied and reviewed%
\cite{Orden98}. It usually requires higher level of quantum chemistry
calculations to determine the fine structures and energy differences between
the isomers. In this paper we focus on providing the lowest-energy
hydrocarbon clusters in various hydrogen chemical potentials and the
lowest-energy pure carbon clusters are intended to be chosen as reference
systems. Nevertheless, the general features (i.e., chain or ring structure)
of the ground state geometries of pure carbon clusters are consistent with
the results in the literature\cite{Orden98,Jones97}.

\subsection{$C_{6}H_{m}$}

\begin{figure}[tbp]
\begin{center}
\includegraphics[width=8.00cm]{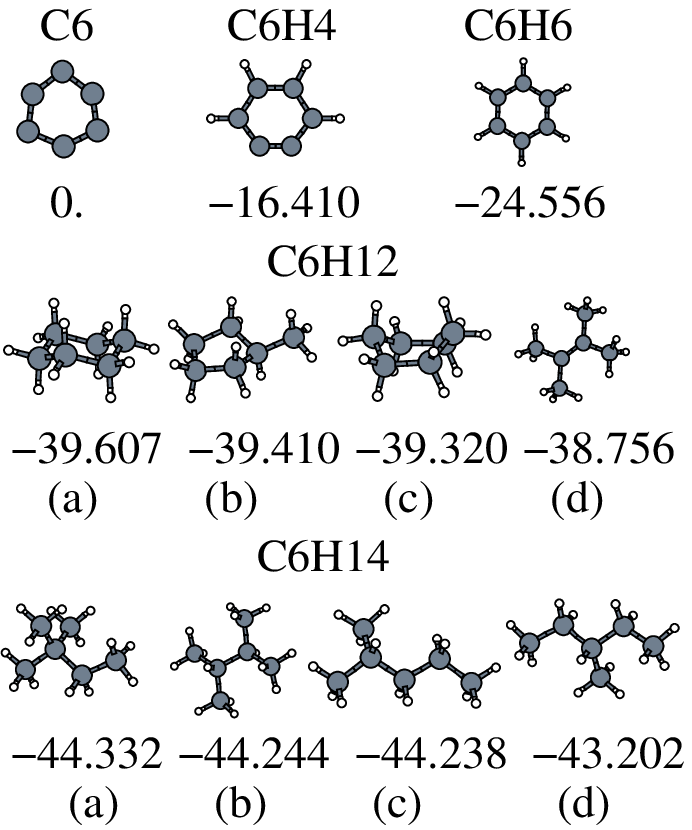} 
\end{center} 
\caption{
11 lowest energy structures of $%
C_{6}H_{m}$ in full hydrogen chemical potential range. The number below each
structure is the relative energy E$_{0}$ in unit of eV.11 lowest energy structures of $%
C_{6}H_{m}$ in full hydrogen chemical potential range. The number below each
structure is the relative energy E$_{0}$ in unit of eV.
}
\label{C6} 
\end{figure} 

Fig. \ref{C6} shows the lowest energy structures with a backbone of 6 carbon
atoms in the full hydrogen chemical potential range. The relative energy E$%
_{0}$ associated with each structure is defined as
\begin{equation}
E_{0}=E_{C_{n}H_{m}}-E_{C_{n}}-mE_{H}
\end{equation}%
where $E_{C_{n}H_{m}}$, $E_{C_{n}}$and $E_{H}$ are the total energies of the
hydrocarbon $C_{n}H_{m}$, pure carbon cluster $C_{n}$ and hydrogen atom $H$,
with $E_{H}=-1.356eV.$ The energies are evaluated at MP2/6-31G(d) level. The
pure $C_{6}$ cluster is chosen as the reference point. In this C6 group, the
lowest-energy hydrocarbon structures in the hydrogen chemical potential
range of $\left( -6.000eV,-4.155eV\right) $, $\left(
-4.155eV,-4.073eV\right) $, $\left( -4.073eV,-2.473eV\right) $, $\left(
-2.473eV,-2.363eV\right) $, $\left( -2.363eV,0\right) $ are $C_{6}$ (chain),
$C_{6}H_{4}$ (Benzyne), $C_{6}H_{6}$ $($Benzene), $C_{6}H_{12}$
(a,Cyclohexane) and $C_{6}H_{14}$ (a,2,2-Dimethylbu$\tan $e), respectively.
Several low energy isomers in $C_{6}H_{12}$ and $C_{6}H_{14}$ group are also
shown for reference. Cyclohexane with boat conformation $C_{6}H_{12}$(c) is $%
0.287eV$ higher in energy than Cyclohexane with chair conformation $%
C_{6}H_{12}$(a), close to the experimental result 0.238 eV\cite{Wiki1}.
Methylcyclopentane $C_{6}H_{12}$(b) is in between and 0.197 eV higher than $%
C_{6}H_{12}$(a), close to experimental value 0.177 eV\cite{Pedleyhandbook}. $%
C_{6}H_{12}$(d) is rather high in energy because of its diradical structure.
The energy orders of the four alkane isomers are correctly predicted by MP2
as compared with experimental results although the energy difference can be
quite small\cite{Pedleyhandbook}.

\subsection{$C_{7}H_{m}$}

\begin{figure}[tbp]
\begin{center}
\includegraphics[width=8.00cm]{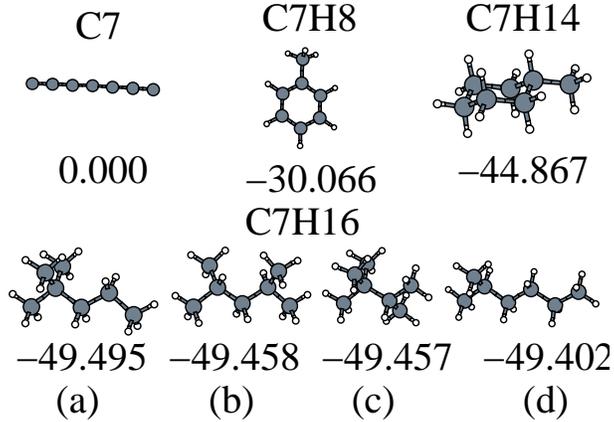}                  
\end{center}  
\caption{
7 lowest energy structures of $%
C_{7}H_{m}$ in full hydrogen chemical potential range. The number below each
structure is the relative energy E$_{0}$ in unit of eV.
} 
\label{C7}     
\end{figure}  

Fig. \ref{C7} shows the lowest energy structures with a backbone of 7 carbon
atoms in the full hydrogen chemical potential range. In this C7 group, the
lowest-energy hydrocarbon structures in hydrogen chemical potential range of
$\left( -6.000eV,-3.758eV\right) $, $\left( -3.758eV,-2.467eV\right) $, $%
\left( -2.467eV,-2.314eV\right) $, $\left( -2.314eV,0\right) $ are $C_{7}$
(chain)$,C_{7}H_{8}$ (Methylbenzene)$,C_{7}H_{14}$ (Methylcyclohexane) and $%
C_{7}H_{16}$ (a,2,2-Dimethylpen$\tan $e), respectively. From C6 to C7 group,
the system mainly choose to add one methyl to the dominant lowest energy
hydrocarbon molecules (i.e., Benzene and Cyclohexane) and exhibit four
different lowest energy chemical compositions in the whole hydrogen chemical
potential range. The alkane subgroup is different and still the one with one
end methyl pair has lowest energy, in consistence with experiment\cite%
{Pedleyhandbook}.

\subsection{$C_{8}H_{m}$}

\begin{figure}[tbp]
\begin{center}
\includegraphics[width=8.00cm]{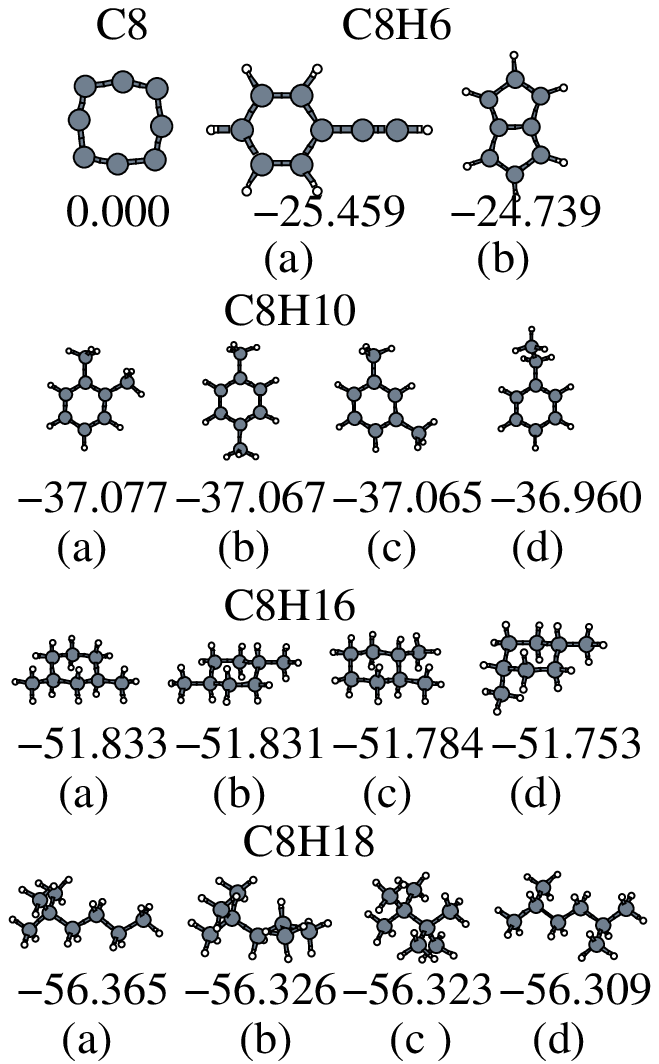}
\end{center}   
\caption{ 
15 lowest energy structures of $%
C_{8}H_{m}$ in full hydrogen chemical potential range. The number below each
structure is the relative energy E$_{0}$ in unit of eV.
}  
\label{C8}      
\end{figure}   

Fig. \ref{C8} shows the lowest energy structures with a backbone of 8 carbon
atoms in the full hydrogen chemical potential range. In this C8 group, the
lowest-energy hydrocarbon structures in hydrogen chemical potential range $%
\left( -6.000eV,-4.243eV\right) $, $\left( -4.243eV,-2.905eV\right) $, $%
\left( -2.905eV,-2.459eV\right) $, $\left( -2.459eV,-2.266\right) $, $\left(
-2.266eV,0\right) $ are $C_{8}$ (ring), $C_{8}H_{6}$ (a,Ethynylbenzene), $%
C_{8}H_{10}$ (a,1,2-Dimethylbenzene), $C_{8}H_{16}$
(a,cis-1,3-Dimethylcyclohexane) and $C_{8}H_{18}$ (a,2,2-Dimethyhexane),\
respectively. Ethynylbenzene has lower energy than $C_{8}H_{6}$(b,Pentalene)
because of its conjugated bond configuration. The first three low energy
Benzene-based isomers in C8H10 subgroup are very close in energy. MP2-level
calculations can not correctly predict the energy orders between them and
favor 1,2-Dimethylbenzene. In contrast, gas phase thermochemistry
measurement shows that $C_{8}H_{10}$ (c,1,3-Dimethylbenzene) is 0.018 eV
lower in energy than 1,2-Dimethylbenzene\cite{Pedleyhandbook}. On the other
hand, MP2 predicts that $C_{8}H_{10}$ (d,Ethylbenzene) is higher in energy
than the Dimethylbenzenes in agreement with experiment\cite{Pedleyhandbook}.
For the C8H16 subgroup, MP2 also predicts that Dimethylcyclohexanes have
lower energies than Ethylcyclohexane and the energy differences between
Dimethylcyclohexanes can be very small. In the alkane subgroup, the one with
one end methyl pair is still the lowest energy isomer.

\subsection{$C_{9}H_{m}$}

\begin{figure}[tbp]
\begin{center}
\includegraphics[width=8.00cm]{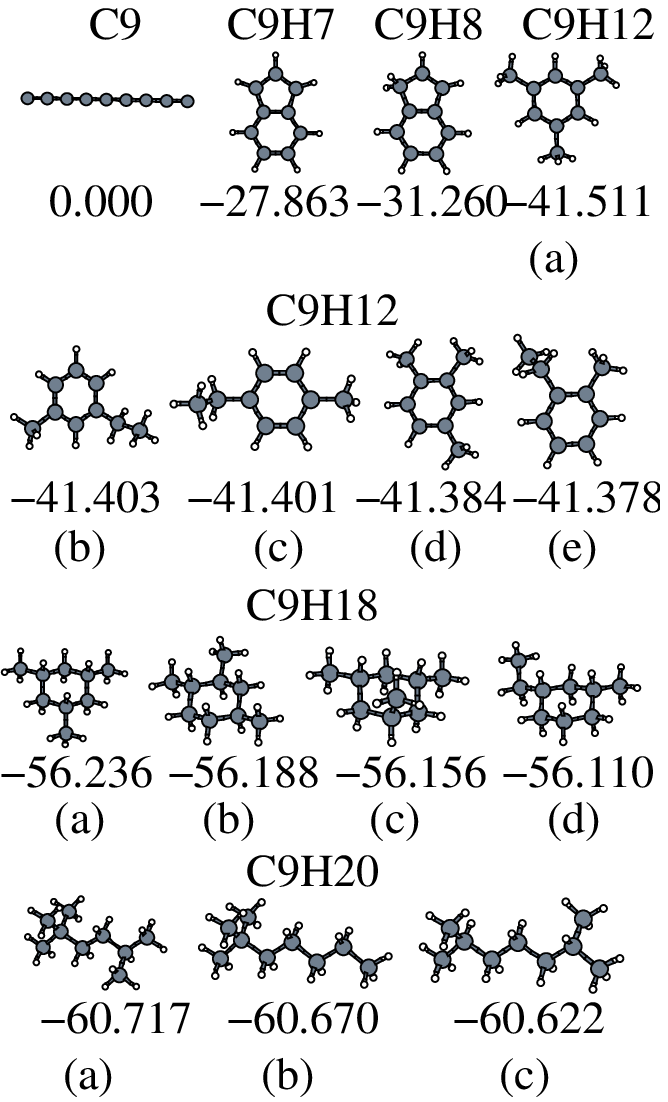}
\end{center}
\caption{
15 lowest energy structures of
$C_{9}H_{m}$ in full hydrogen chemical potential range. The number below
each structure is the relative energy E$_{0}$ in unit of eV.
}   
\label{C9}       
\end{figure}    

Fig. \ref{C9} shows the lowest energy
structures with a backbone of 9 carbon atoms in the full hydrogen chemical
potential range. In this C9 group, the lowest-energy hydrocarbon structures
in hydrogen chemical potential range $\left( -6.000eV,-3.981eV\right) $, $%
\left( -3.981eV,-3.397eV\right) $, $\left( -3.397eV,-2.563eV\right) $, $%
\left( -2.563eV,-2.454eV\right) $, $\left( -2.454eV,-2.240eV\right) $, $%
\left( -2.20eV,0\right) $ are $C_{9}$ (chain)$,C_{9}H_{7}$, $C_{9}H_{8}$
(indene), $C_{9}H_{12}$ (a,1,2,5-Trimethylbenzene), $C_{9}H_{18}$
(1,3,5-Trimethylcyclohexane) and $C_{9}H_{20}$ (a,2,2,5-Trimethyhexane),\
respectively. The group shows more ground state chemical compositions in the
full hydrogen chemical potential range. MP2 predicts that the isomer with
1,3,5-Trimethyl has the lowest energy in both C9H12 and C9H18 subgroups due
to maximal reduction of the repulsion energy between the hydrogen atoms. The
ground state geometry of the alkane subgroup is predicted to have three
methyls in accordance with experiment\cite{Pedleyhandbook}.

\subsection{$C_{10}H_{m}$}

\begin{figure}[tbp]
\begin{center}
\includegraphics[width=8.00cm]{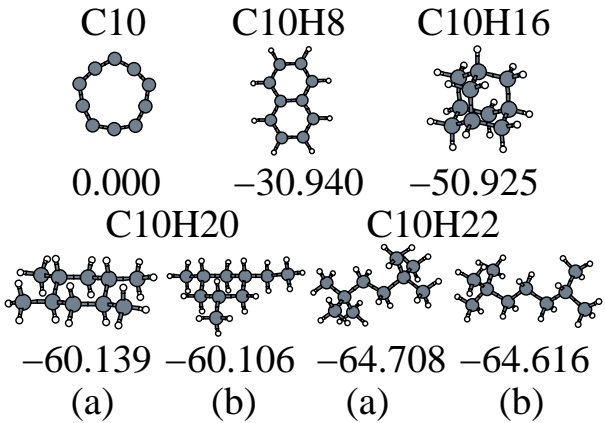}
\end{center} 
\caption{ 
7 lowest energy structures of $%
C_{10}H_{m}$ in full hydrogen chemical potential range. The number below
each structure is the relative energy E$_{0}$ in unit of eV.
}    
\label{C10}        
\end{figure}   

Fig. \ref{C10} shows the lowest energy
structures with a backbone of 10 carbon atoms in the full hydrogen chemical
potential range. In this C10 group, the lowest-energy hydrocarbon structures
in hydrogen chemical potential range $\left( -6.000eV,-3.868eV\right) $, $%
\left( -3.868eV,-2.498eV\right) $, $\left( -2.498eV,-2.303eV\right) $, $%
\left( -2.303eV,-2.285eV\right) $, $\left( -2.285eV,0\right) $ are $C_{10}$
(ring), $C_{10}H_{8}$ (Naphthalene), $C_{10}H_{16}$ (Adaman$\tan $e), $%
C_{10}H_{20}$ (a,trans-1,2,4,5-Tetramethylcyclohexane) and $C_{10}H_{22}$
(a,2,2,5,5-Tetramethyhexane),\ respectively. Interestingly, the smallest 3D
fragment of diamond (i.e., Adaman$\tan $e) appears as a ground state
geometry in the group. The corresponding chemical potential range is quite
narrow though. The 2D cyclohexane-based structure is still present. The
isomer with methy pair at both ends is predicted to have the ground state
geometry in the alkane subgroup.

\subsection{$C_{11}H_{m}$}

\begin{figure}[tbp]
\begin{center} 
\includegraphics[width=8.00cm]{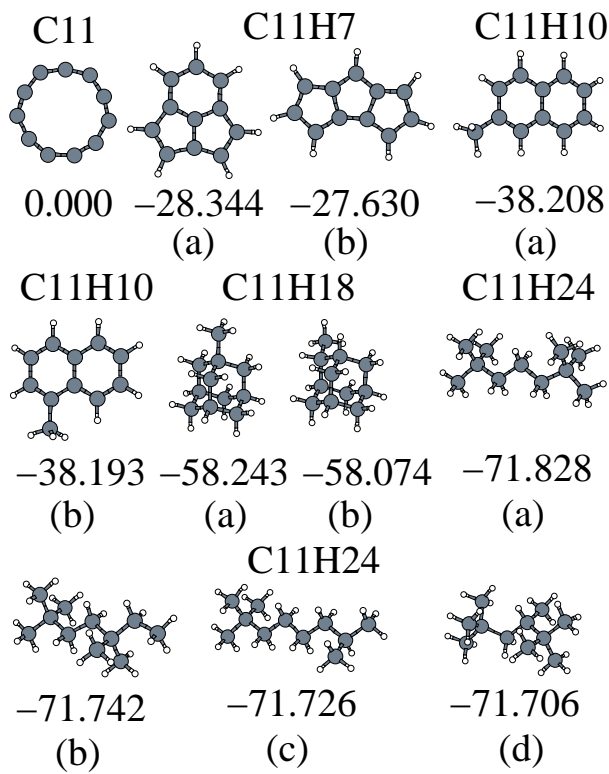}
\end{center}  
\caption{  
11 lowest energy structures of $%
C_{11}H_{m}$ in full hydrogen chemical potential range. The number below
each structure is the relative energy E$_{0}$ in unit of eV.
}     
\label{C11}         
\end{figure}    

Fig. \ref{C11} shows the lowest energy
structures with a backbone of 11 carbon atoms in the full hydrogen chemical
potential range. In this C11 group, the lowest-energy hydrocarbon structures
in hydrogen chemical potential range $\left( -6.000eV,-4.049eV\right) $, $%
\left( -4.049eV,-3.288eV\right) $, $\left( -3.288eV,-2.504eV\right) $, $%
\left( -2.504eV,-2.264eV\right) $, $\left( -2.264eV,0\right) $ are $C_{11}$
(ring), $C_{11}H_{7}$, $C_{11}H_{10}$ (a,2-Methy$\ln $aphthalene), $%
C_{11}H_{18}$ (a,1-Methyladaman$\tan $e) and $C_{11}H_{24}$
(a,2,2,5,5-Tetramethyhep$\tan $e),\ respectively. The energy order of the
isomers in C11H10 subgroup is correctly predicted by MP2 compared with
experiment although the energy difference is tiny\cite{Pedleyhandbook}. MP2
favors the growth of one methyl at the monohydrogen site and predicts that
1-Methyladaman$\tan $e is 0.169 eV lower in energy than $C_{11}H_{18}$
(b,2-Methyladaman$\tan $e), close to the experimental value of 0.206 eV\cite%
{Pedleyhandbook}. One may also observe that the 2D cyclohexane-based
structure disappears in the ground state geometries of the hydrocarbons with
carbon number larger than ten. Thus we observe a crossover between
populations of 2D cyclohexane-based structure and 3D diamond fragment with
increasing size of hydrocarbon. The ground state isomer in the alkane
subgroup is still predicted to the one with methyl pair at both ends.

\subsection{$C_{12}H_{m}$}

\begin{figure}[tbp]
\begin{center}
\includegraphics[width=8.00cm]{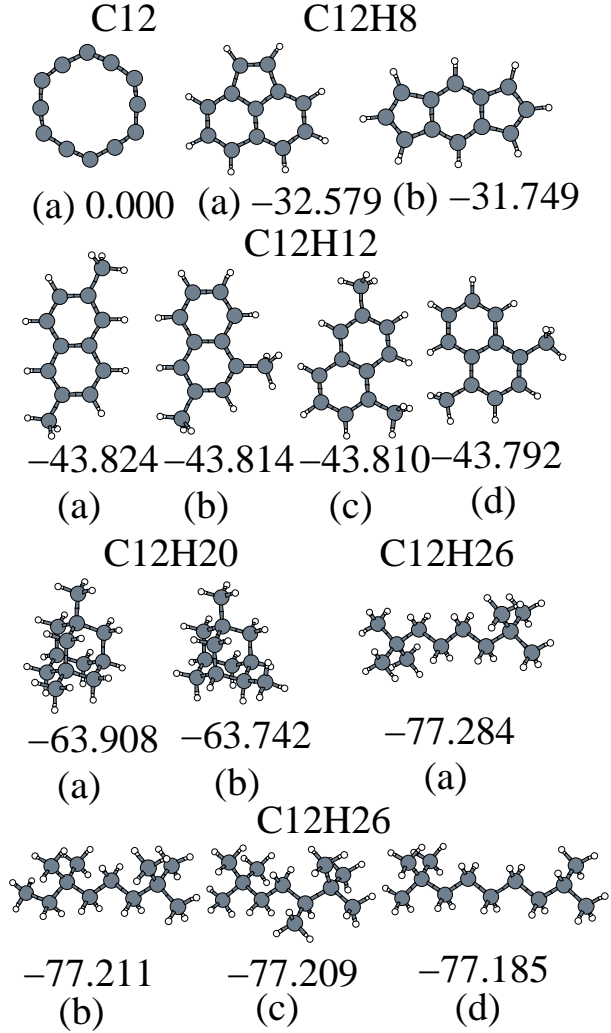}
\end{center}   
\caption{  
13 lowest energy structures of
$C_{12}H_{m}$ in full hydrogen chemical potential range. The number below
each structure is the relative energy E$_{0}$ in unit of eV.
}      
\label{C12}        
\end{figure}    

Fig. \ref{C12} shows the lowest energy
structures with a backbone of 12 carbon atoms in the full hydrogen chemical
potential range. In this C12 group, the lowest-energy hydrocarbon structures
in hydrogen chemical potential range $\left( -6.000eV,-4.198eV\right) $, $%
\left( -4.198eV,-2.561eV\right) $, $\left( -2.561eV,-2.511eV\right) $, $%
\left( -2.511eV,-2.230eV\right) $, $\left( -2.230eV,0\right) $ are $C_{12}$
(ring), $C_{12}H_{8}$ (Acenaphthylene), $C_{12}H_{12}$ (a,2,6-Dimethy$\ln $%
aphthalene), $C_{12}H_{20}$ (a,1,3-Dimethyladaman$\tan $e) and $C_{12}H_{26}$
(a,2,2,7,7-Tetramethyloc$\tan $e),\ respectively. Similar rules as mentioned
above can be applied here: 2,6-Dimethy$\ln $aphthalene is the lowest energy
isomer because of the maximal reduction of the repulsion energy between the
hydrogen atoms; monohydrogen site is the preferred site to grow methyl on
adamantane; the isomer with methyl pair at both ends is still lowest in
energy in the alkane subgroup.

\subsection{$C_{13}H_{m}$}

\begin{figure}[tbp]
\begin{center}
\includegraphics[width=8.00cm]{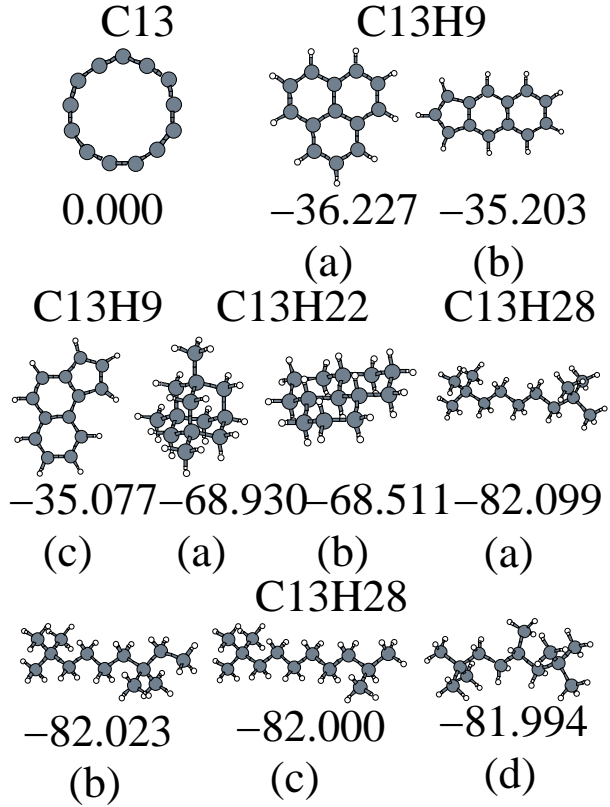}
\end{center}    
\caption{
10 lowest energy structures of $%
C_{13}H_{m}$ in full hydrogen chemical potential range. The number below
each structure is the relative energy E$_{0}$ in unit of eV.10 lowest energy structures of $%
C_{13}H_{m}$ in full hydrogen chemical potential range. The number below
each structure is the relative energy E$_{0}$ in unit of eV.
}      
\label{C13}        
\end{figure}    

Fig. \ref{C13} shows the lowest energy
structures with a backbone of 13 carbon atoms in the full hydrogen chemical
potential range. In this C13 group, the lowest-energy hydrocarbon structures
in hydrogen chemical potential range $\left( -6.000eV,-4.025eV\right) $, $%
\left( -4.025eV,-2.516eV\right) $, $\left( -2.516eV,-2.195eV\right) $, $%
\left( -2.195eV,0\right) $ are $C_{13}$ (ring), $C_{13}H_{9}$ (Phenalene), $%
C_{13}H_{22}$ (a,1,3,5-Trimethy$\ln $aphthalene) and $C_{13}H_{28}$
(a,2,2,8,8-Tetramethy$\ln $onane), respectively. Phenalene is the lowest
energy isomer in C13H9 subgroup due to its closely packed aromatic rings.
1,3,5-Trimethy$\ln $aphthalene is predicted to be much lower in energy than $%
C_{13}H_{22}$(b,dodecahydro-1H-Phenalene). The ground state structure in the
alkane subgroup still keeps the single carbon chain with methyl pair at both
ends.

\subsection{$C_{14}H_{m}$}

\begin{figure}[tbp]
\begin{center}
\includegraphics[width=8.00cm]{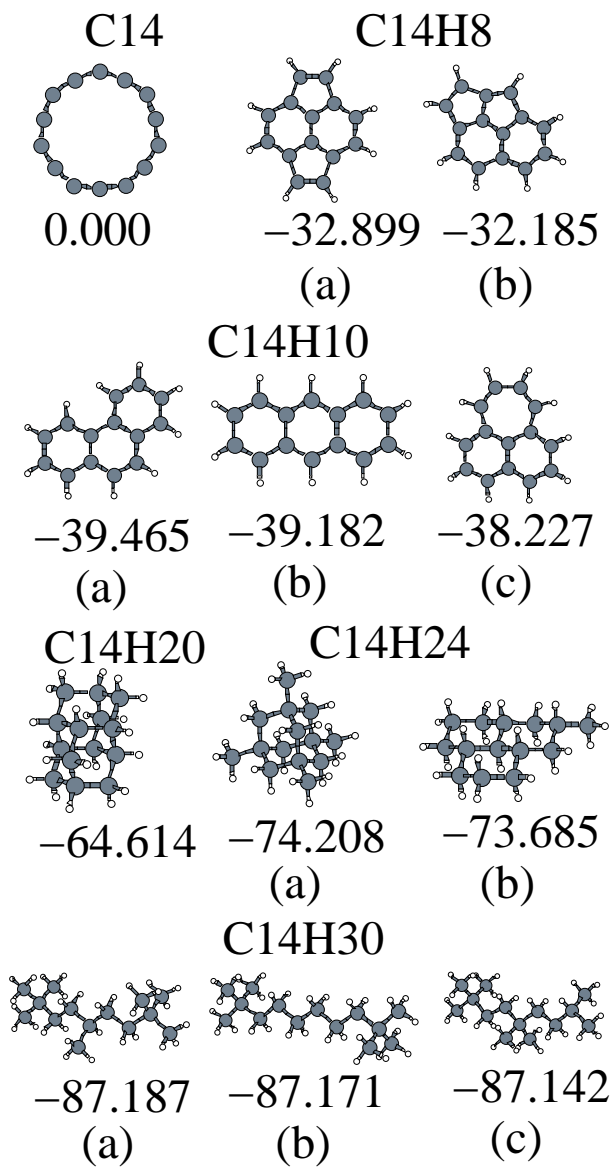}
\end{center}     
\caption{ 
12 lowest energy structures of
$C_{14}H_{m}$ in full hydrogen chemical potential range. The number below
each structure is the relative energy E$_{0}$ in unit of eV.
}       
\label{C14}
\end{figure}

Fig. \ref{C14} shows the lowest energy
structures with a backbone of 14 carbon atoms in the full hydrogen chemical
potential range. In this C14 group, the lowest-energy hydrocarbon structures
in hydrogen chemical potential range of $\left( -6.000eV,-4.113eV\right) $, $%
\left( -4.113eV,-3.283eV\right) $, $\left( -3.283eV,-2.515eV\right) $, $%
\left( -2.515eV,-2.399eV\right) $, $\left( -2.399eV,-2.163eV\right) $, $%
\left( -2.163eV,0\right) $ are $C_{14}$ (ring), $C_{14}H_{8}$
(Cyclopent[fg]acenaphthylene), $C_{14}H_{10}$ (a,Phenanthrene), $%
C_{14}H_{20} $ (a,Diaman$\tan $e), $C_{14}H_{24}$
(a,1,3,5,7-Tetramethyladamantane) and $C_{14}H_{30}$ (a,2,2,5,8,8-penmethy$%
\ln $onane), respectively. Contrary to the rule of maximal reduction of
repulsion energy between hydrogen atoms, Phenanthrene is correctly predicted
to be 0.283 eV lower in energy than $C_{14}H_{10}$(b,Anthracene), comparable
to the experimental result of 0.216 eV. A bigger diamond fragment (Diaman$%
\tan $e) appears as ground state structure in a narrow hydrogen chemical
potential range. The fifth methyl appears in the middle of the carbon chain
besides methyl pair at both ends of the lowest energy isomer in the alkane
subgroup.

\subsection{$C_{15}H_{m}$}

\begin{figure}[tbp]
\begin{center}
\includegraphics[width=8.00cm]{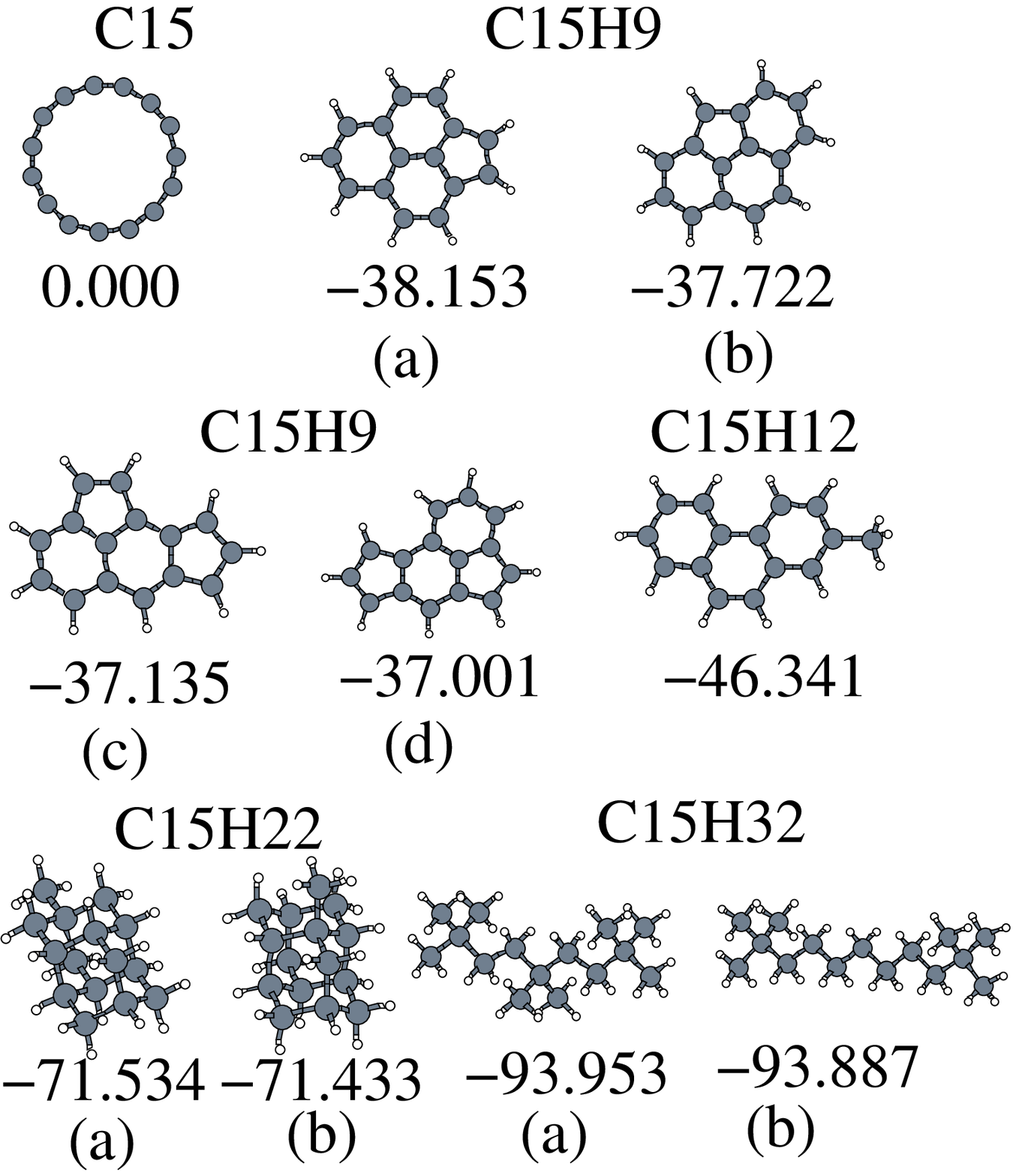}
\end{center}      
\caption{  
10 lowest energy structures of $%
C_{15}H_{m}$ in full hydrogen chemical potential range. The number below
each structure is the relative energy E$_{0}$ in unit of eV.
}        
\label{C15}
\end{figure}

Fig. \ref{C15} shows the lowest energy
structures with a backbone of 15 carbon atoms in the full hydrogen chemical
potential range. In this C15 group, the lowest-energy hydrocarbon structures
in hydrogen chemical potential range of $\left( -6.000eV,-4.239eV\right) $, $%
\left( -4.239eV,-2.729eV\right) $, $\left( -2.720eV,-2.519eV\right) $, $%
\left( -2.519eV,-2.243eV\right) $, $\left( -2.243eV,0\right) $ are $C_{15}$
(ring), $C_{15}H_{9}$ (a), $C_{15}H_{12}$ (3-Methylphenanthrene), $%
C_{15}H_{22}$ (a,3-Methyldiadamantane) and $C_{15}H_{32}$
(a,2,2,5,5,8,8-Hexamethy$\ln $onane), respectively. $C_{15}H_{9}$ (a) is
predicted to be the lowest energy isomer due to its compact structure and
minimal influence of the five fold ring. Methyl is still preferred to grow
at the monohydrogen site of diadamantane. Three pairs of methyls are
distributed symmetrically on the carbon chain in the ground state structure
of alkane subgroup.

\subsection{$C_{16}H_{m}$}

\begin{figure}[tbp]
\begin{center}
\includegraphics[width=8.00cm]{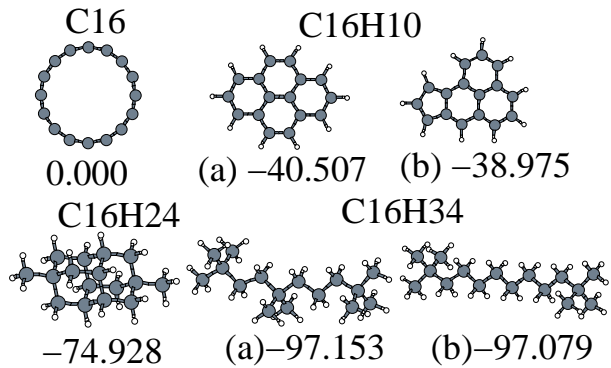}
\end{center}       
\caption{  
6 lowest energy structures of $%
C_{16}H_{m}$ in full hydrogen chemical potential range. The number below
each structure is the relative energy E$_{0}$ in unit of eV.
}         
\label{C16}
\end{figure}

Fig. \ref{C16} shows the lowest energy
structures with a backbone of 16 carbon atoms in the full hydrogen chemical
potential range. In this C16 group, the lowest-energy hydrocarbon structures
in hydrogen chemical potential range $\left( -6.000eV,-4.051eV\right) $, $%
\left( -4.051eV,-2.459eV\right) $, $\left( -2.459eV,-2.223eV\right) $, $%
\left( -2.223eV,0\right) $ are $C_{16}$ (ring), $C_{16}H_{10}$ (a,Pyrene), $%
C_{16}H_{24},$ and $C_{16}H_{34}$ (a,2,2,5,5,9,9-Hexamethyloc$\tan $e),
respectively.

\subsection{$C_{17}H_{m}$}

\begin{figure}[tbp]
\begin{center}
\includegraphics[width=8.00cm]{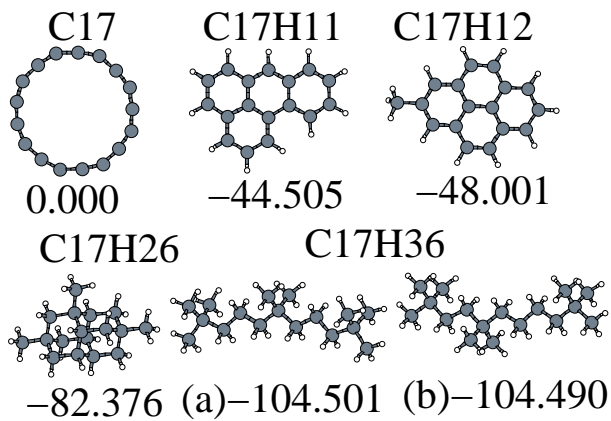}
\end{center}        
\caption{
6 lowest energy structures of $%
C_{17}H_{m}$ in full hydrogen chemical potential range. The number below
each structure is the relative energy E$_{0}$ in unit of eV.
}         
\label{C17}
\end{figure}

Fig. \ref{C17} shows the lowest energy
structures with a backbone of 17 carbon atoms in the full hydrogen chemical
potential range. In this C17 group, the lowest-energy hydrocarbon structures
in hydrogen chemical potential range of $\left( -6.000eV,-4.046eV\right) $, $%
\left( -4.046eV,-3.496eV\right) $, $\left( -3.496eV,-2.456eV\right) $, $%
\left( -2.456eV,-2.213eV\right) $, $\left( -2.213eV,0\right) $ are $C_{17}$
(ring), $C_{17}H_{11}$, $C_{17}H_{12}$ (2-Methylpyrene), $C_{17}H_{26}$, and
$C_{17}H_{36}$ (a,2,2,6,6,10,10-Hexamethyludecane), respectively.

\subsection{$C_{18}H_{m}$}

\begin{figure}[tbp]
\begin{center}
\includegraphics[width=8.00cm]{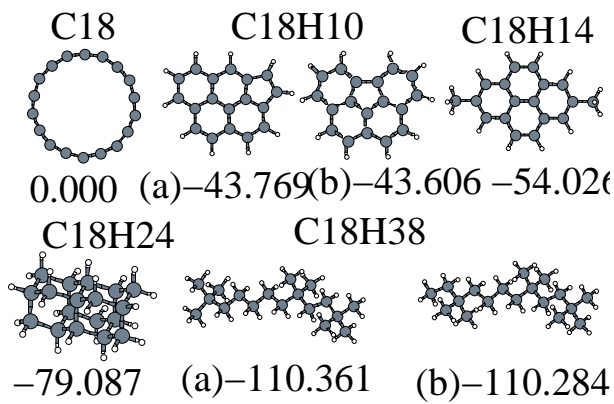}
\end{center}         
\caption{ 
7 lowest energy structures of $%
C_{18}H_{m}$ in full hydrogen chemical potential range. The number below
each structure is the relative energy E$_{0}$ in unit of eV.
}          
\label{C18}
\end{figure}

Fig. \ref{C18} shows the lowest energy
structures with a backbone of 18 carbon atoms in the full hydrogen chemical
potential range. In this C18 group, the lowest-energy hydrocarbon structures
in hydrogen chemical potential range $\left( -6.000eV,-4.377eV\right) $, $%
\left( -4.377eV,-2.564eV\right) $, $\left( -2.564eV,-2.506eV\right) $, $%
\left( -2.506eV,-2.234eV\right) $, $\left( -2.234eV,0\right) $ are $C_{18}$
(ring), $C_{18}H_{10}\left( a\right) $, $C_{18}H_{14}$ (2,7-Dimethylpyrene),
$C_{18}H_{24}$ (Diamantane) and $C_{18}H_{38}$
(a,2,2,5,5,11,11-Hexamethyldodecane), respectively. Similar empirical rules
are observed in C16-C18 groups: Poly-aromatic rings tend to form compact
structure; Methyl favors the monohydrogen site in diamond fragment
structure; ground state structure of alkane subgroup favors methyl pairs
spaced by at least two carbon atoms over the main carbon chain.

\section{Conclusion}

An unbiased evolution-based optimization method combined with Brenner's
empirical potential is used to search for ground state structures of
hydrocarbon molecules in a wide hydrogen chemical potential range. The
resultant structures are further sorted by quantum chemical calculations at
MP2 level. The collection of lowest energy structures of hydrocarbon
molecules $C_{m}H_{n}$ (m=1-18; n=0-2m+2) is presented. A crossover between
populations of 2D cyclohexane-based structure and 3D diamond fragment with
increasing number of carbon atoms of the hydrocarbon molecules is
demonstrated. Besides the PAH\ compounds, we also show that PAH with methyls
can also be important in the interstellar medium. The spectra of the PAH
compounds and how are they affected by the various configurations of the
radicals (e.g., methyl) are worthy of further studying.

\section{Acknowledgments}

We want to thank M.S. Tang for many useful discussions. Work at the Ames
laboratory was supported by the U.S. Department of Energy, Office of Basic
Energy Science, Division of Materials Science and Engineering including a
grant of computer time at the National Energy Research Supercomputing Center
(NERSC) at the Lawrence Berkeley National Laboratory under Contract No.
DE-AC02-07CH11358. C. Rareshide acknowledges the support from NSF\ sponsored
Research Experience for Undergraduates (REU) program at Iowa State
University.

\end{document}